\documentclass[letterpaper, 10 pt, conference]{ieeeconf}

\IEEEoverridecommandlockouts                  
\usepackage{cite}
\usepackage{amsmath,amssymb,amsfonts,bm}
\usepackage{enumerate}
\usepackage{graphicx}
\usepackage{textcomp}
\usepackage{xcolor}
\usepackage{color}
\usepackage{amsthm}
\usepackage{amsmath}

\usepackage[labelsep=period]{caption}

\usepackage{tikz}
\usepackage{dsfont}
\newtheorem{thm}{Theorem}

\newtheorem{lem}[thm]{Lemma}

\newtheorem{prot}{Protocol}
\theoremstyle{definition}

\newtheorem{defn}{Definition}
\newtheorem{rem}{Remark}

\newtheorem{prob}{Problem}
\usepackage[utf8]{inputenc}
\usepackage{color}

\usepackage[titles]{tocloft}

\usepackage{hyperref}

\begin{document}

\title{Linear Regulator-Based Synchronization of Positive Multi-Agent Systems}

\author{Alba Gurpegui, Mark Jeeninga, Emma Tegling, Anders Rantzer
\thanks{The authors are with the Department of Automatic Control and the ELLIIT Strategic Research Area at Lund University, Lund, Sweden. Email: \href{mailto:alba.gurpegui@control.lth.se}{alba.gurpegui@control.lth.se}, 
\href{mailto:emma.tegling@control.lth.se}{mark.jeeninga@control.lth.se},
\href{mailto:emma.tegling@control.lth.se}{emma.tegling@control.lth.se}, 
\href{mailto:anders.rantzer@control.lth.se}{anders.rantzer@control.lth.se}. }
\thanks{This work is partially funded by the Wallenberg AI, Autonomous Systems and Software Program (WASP) and the European Research Council (ERC) under the European Union's Horizon 2020 research and innovation programme under grant agreement No 834142  (ScalableControl).}}
\maketitle
\begin{abstract}
This paper addresses the positive synchronization of interconnected systems on undirected graphs. For homogeneous positive systems, a static feedback protocol design is proposed, based on the Linear Regulator problem. The solution to the algebraic equation associated to the stabilizing policy can be found using a linear program. Necessary and sufficient conditions on the positivity of each agent's trajectory for all nonnegative initial conditions are also provided. Simulations on large regular graphs with different nodal degree illustrate the proposed results.
\end{abstract}
\section{Introduction}
\subsection{Motivation}
A remarkable problem in control theory is the design of protocols that lead to the synchronization of interconnected systems.  Synchronization is a desired behavior in many dynamical systems associated with numerous applications~\cite{18, Emma_scale_frag, 9, 10}. 
This paper considers multi-agent systems with homogeneous, linear time-invariant dynamics, and is concerned with the synchronization of the states of such agents using only relative measurements.

Over the last decade, several works have addressed this synchronizing problem.
A particularly popular approach is based on finding a stabilizing controller for the closed-loop agent dynamics, which then can be used to formulate a controller that achieves synchronization in the overall system.
The textbook \cite{book_ref} gives a detailed overview of the general approach.
A key tool in stabilizing the local system is to solve the algebraic Riccati equation for the agent dynamics.
For the resulting controller it holds that increasing the input signal will also still stabilize the local system, which is a prerequisite for achieving synchronization of the multi-agent system. 

An important characteristic of our work is the positive dynamics. Positive systems are characterized by the property that their state and output remain nonnegative for any nonnegative input and initial state. Classical references on this topic are~\cite{BermanBook} and~\cite{Luenberger}.
Extensive research has been conducted into expanding the scope of positive systems theory, exploring its natural extensions.  For instance, past literature has attempted to solve the positive consensus problem \cite{positive_consensus1, positive_consensus2}. In particular, in~\cite{pos_sys_MAS_imp_} this problem is solved implementing a static output feedback approach. This method is designed for positive agents, with the controller gain matrix structured to represent the network connectivity.

This paper also formulates a controller that achieves synchronization in the case that the agent has positive homogeneous dynamics. The approach is based on solving the Linear Regulator problem \cite{LR_paperII}, which is analogous to the algebraic Riccati equation in the Linear Quadratic Regulator problem, but has the advantage that its associated algebraic equation can be solved using a linear program. Moreover, while the Riccati equation and the number of unknown parameters grows quadratically with the state dimension, in the Linear Regulator setting, the algebraic equation growth rate is linear with respect to the dimension of the state.

The main contributions of the paper are: (i) A static feedback protocol based on the algebraic equation of the Linear Regulator problem, which can be solved with linear programming (Protocol~\ref{protocol_LR}). (ii) Sufficient and necessary conditions on the positivity of each agent's trajectory for all nonnegative initial conditions (Theorem~\ref{thm_nonneg}). (iii) Analysis on the eigenvalue bounds of the graph families that are relevant to this manuscript (Section~\ref{sec/graphs}).

\section{Model Description}
\subsection{Notation}
Let $\mathbb{R}_{+}$ denote the set of nonnegative real numbers, $\mathbb{R}^{n}$ the $n$-dimensional Euclidean space and $\mathbb{R}^{n}_{+}$ the positive orthant. $\mathbb{R}^{m \times n}$ denote the set of $m$ by $n$ matrices. Any vector is, by default, a column vector. $\mathds{1}$ denotes a column vector with unit entries in $\mathbb{R}$ of appropriate dimension. $I_n$ denotes the identity matrix of dimension $n \times n$. For a real matrix $X$ $\sigma(X)$ denotes the spectrum, $\rho(X)$ the spectral radius and $\left | X \right |$ denotes the matrix obtained by replacing the elements of $X$ with their absolute values. A real matrix $X$ is called nonnegative $X \geq 0$ (respectively nonpositive $X\leq0)$ if all the elements of $X$ are nonnegative (nonpositive) but at least one element is nonzero. If all the elements of $X$ are strictly positive (strictly negative), we call $X$ a positive (negative) matrix. For real matrices $X,Y$, the inequality $X \geq Y$ $(X \leq Y)$ means that all the elements of the matrix $X-Y$ are nonnegative (nonpositive). Because column and row vectors are special forms of matrices, the function $\left | \cdot  \right |$ and relations $\geq$, $\leq$, $>$, $<$ apply to them. Lastly, let $\otimes$ denote the Kronecker product.
\subsection{Graph Description}
A graph $\mathcal{G}$ is also defined as a set $\mathcal{G}=\left\{ \mathcal{V}_{\mathcal{G}}, \mathcal{E}_{\mathcal{G}} \right\}$, with a set of $N= \left| \mathcal{V}_{\mathcal{G}} \right|$ nodes and a set $\mathcal{E}_{\mathcal{G}}\subset \mathcal{V}_{\mathcal{G}} \times \mathcal{V}_{\mathcal{G}}$ of edges, each of which has an associated nonnegative weight $w_{ij}$. Since only undirected graphs are considered in this work, the edge $(i,j)\in \mathcal{E}_{\mathcal{G}}$ is bidirectional (i.e. it does not have a direction). 
Moreover, every graph $\mathcal{G}$ has a connected  spanning tree. This means there exists a path from any node $i \in \mathcal{V}_{\mathcal{G}}$ to every other node $j \in \mathcal{V}_{\mathcal{G}} \setminus { i}$.

The Laplacian matrix $\mathcal{L}(\mathcal{G})$ of $\mathcal{G}$, or simply $\mathcal{L}$ if $\mathcal{G}$ is clear from the context, is defined as 
\[  \mathcal{L}(\mathcal{G})_{ij}= \left\{
    \begin{array}{ll}
      -w_{ij} & i\neq j \\
      \sum_{k \in \mathcal{N}_i}w_{ik} & i=j \\
      0 & \mathrm{otherwise}. 
\end{array} 
\right. \]
The Laplacian matrix of undirected graphs is symmetric, thus, its eigenvalues are real.

The set of connected graphs is denoted by $\mathbb{G}$ and is characterized by the following \nobreak{property.}
\begin{lem}\label{lem_conn_eig0_eiv1}
    A graph $\mathcal{G}$ is connected if and only if the associated Laplacian matrix $\mathcal{L}$ has a simple eigenvalue at the origin. Furthermore, in this case, the eigenvector associated with the eigenvalue at the origin is $\mathds{1}$ and all other eigenvalues lie in the open right half-plane.
\end{lem}
Denote by $\lambda_k(\mathcal{G})$, or $\lambda_k$ if $\mathcal{G}$ is clear from the context, with $k=1,\dots N$ the eigenvalues of $\mathcal{L}$. The eigenvalues are numbered so that $0=\lambda_1 < \lambda_2\leq \dots \leq \lambda_N$. The eigenvalue~$\lambda_2$ is known as the algebraic connectivity of $\mathcal{G}$.

This paper considers graph families $\mathcal{F}\subseteq \mathbb{G}$ within the set of connected undirected graphs. Moreover, an upper and lower bound for the eigenvalues of the Laplacian matrix associated to the graphs are assumed. 
\begin{defn}\label{G_alphabeta} The set of undirected graphs for which the associated Laplacian matrix has nonzero eigenvalues $\lambda_i$, $i=2, \dots N$ such that $\beta \leq \lambda_i \leq \gamma$ is defined by
\begin{align*}
    \mathbb{G}_{[\beta, \gamma]}= \left\{\mathcal{G}\in \mathbb{G} \hspace{1mm} | \hspace{1mm} \lambda_i(\mathcal{G})\in \left[ \beta, \gamma \right], \hspace{2mm} \forall i > 1 \right\}.
\end{align*}
\end{defn}
Recall that $\beta$ is a lower bound for the algebraic connectivity of the graphs in $\mathbb{G}_{[\beta, \gamma]}$.
In the literature, the upper bound $\gamma$ has been be related to the number of agents and their degree of connectivity. 
\subsection{Multi-agent Systems}
We consider a multi-agent system (MAS) composed by an arbitrary number of identical, linear time-invariant agents of the form
\begin{align}\label{MAS_2.1}
    \dot{x}_i&=Ax_i + Bu_i
\end{align}
where $x_i \in \mathbb{R}^n$, $u_i \in \mathbb{R}^m$ are, respectively, the state and input vectors of agent $i$, $A\in \mathbb{R}^{n \times n}$ is Metzler and $B\in \mathbb{R}^{n\times m}$. 

It is assumed that the communication network provides each agent with a linear combination of its own state relative to the states of its neighboring agents. Each agent $i \in \left\{1, \dots, N \right\}$ in the network can access the relative information of the full states of its neighboring agents compared to its own state. Specifically, each agent has access to the quantity
\begin{align}\label{zeta_aij}
    \zeta_i = \sum_{j=1}^N w_{ij}(x_i - x_j) 
\end{align}
where $w_{ij}\geq 0$, $w_{ii}=0$ for $i,j \in \left\{1, \dots, N \right\}$. The network topology is described by an undirected graph $\mathcal{G}\in \mathbb{G}_{[\beta, \gamma]}$ where nodes represent the agents and edges correspond to the nonzero coefficients $w_{ij}$. In particular, $w_{ij}>0$ indicates the presence of an edge from agent $j$ to $i$ with the edge weight equal to the magnitude of $a_{ij}$. 

The communication in the continuous-time network is expressed using the Laplacian matrix $\mathcal{L}$, associated with this weighted graph $\mathcal{G}$. In particular, $\zeta_i$ can be written as 
\begin{align}\label{zeta_lij}
    \zeta_i=\sum_{j=1}^N l_{ij}x_j, \hspace{4mm} \mathcal{L}=(l_{ij}).
\end{align}
When working with multiagent systems (MAS), it is often the case that agents do not know the network graph or its associated weights. 
If $\mathcal{F}$ contains only a single graph, the agents know the graph structure, and protocol~\eqref{protocol_gral} can be designed based on a known Laplacian matrix. However, if $\mathcal{F}$ is large then the protocol must be designed with minimal information about the graph and its Laplacian matrix. Note that, even though lower and upper bounds for the Laplacian eigenvalues are required, this approach still accommodates nearly all possible graphs, as these bounds can be arbitrarily small or large, respectively.
\subsection{Positive Systems}
Positive linear dynamical systems are often characterized through Metzler matrices.
\begin{defn}[Metzler]
    A square matrix $A \in \mathbb{R}^{n \times n}$ is Metzler if its off-diagonal entries are all nonnegative.
\end{defn}
\noindent In general terms, positive systems are characterized by the property that their state and output remain nonnegative for any nonnegative input and initial state. 
\begin{defn}[Positive System]
    A linear system~\eqref{MAS_2.1} is called (internally) positive if and only if its state and output are non-negative for every non-negative input and every non-negative initial state.
\end{defn}
\noindent This type of dynamic behavior has attracted significant interest in control theory due to its capacity to model a wide range of physical phenomena. Foundational works on positive systems include~\cite{BermanBook} and~\cite{Luenberger}.
\begin{thm}[\cite{Luenberger}]
    The continuous-time linear system~\eqref{MAS_2.1} is positive if and only if $A$ is a Metzler matrix and $B\geq 0$.
\end{thm}
Recent advancements in the control of positive systems are surveyed in~\cite{tutorial}. One key advantage of positive systems, their stability can be easily verified using linear Lyapunov functions~\cite{Blanchini_lyapunov}.
\section{Preliminaries}
A multi-agent system is considered homogeneous when all agents have identical dynamics. In such systems, synchronization is achieved when the state differences between agents in the network converge to zero. 
\begin{defn}[State synchronization]
Consider the MAS described by~\eqref{MAS_2.1} and~\eqref{zeta_aij}. The agents in the network achieve state synchronization if:    \begin{align}\label{sync_lim_2.4}
        \lim_{t\rightarrow \infty} \left[ x_i(t)- x_j(t) \right]=0, \hspace{5mm} \forall i, j \in \left\{ 1, \dots, N \right\}
    \end{align}
\end{defn}
\begin{rem}
    The agents in~\eqref{MAS_2.1} are described by an LTI system where $y(t)=x(t)$. This is referred to in the literature as full-state coupling.
\end{rem}
\begin{prob}[State synchronization]\label{problem_gral}
    Consider a MAS described by~\eqref{MAS_2.1},~\eqref{zeta_aij}. Let $\mathcal{F}$ be a given family of graphs such that $\mathcal{F}\subseteq \mathbb{G}$. The state synchronization problem with a set of network graphs $\mathcal{F}$ is to find, if possible, a linear static protocol of the form:
    \begin{align}\label{protocol_gral}
        u_i=F\zeta_i
    \end{align}
    for $i=1, \dots, N$ such that for any graph $\mathcal{G} \in \mathcal{F}$ and for all the initial conditions of agents, the state synchronization among agents is achieved. Furthermore, the problem is referred to as \textit{the positive consensus problem} if, for any selection of nonnegative initial conditions, the states of all agents remain nonnegative.
\end{prob}
After implementing the linear static protocol~\eqref{protocol_gral}, the MAS described by~\eqref{MAS_2.1} and~\eqref{zeta_aij} follows from the dynamics
\begin{align*}
    \dot{x}_i=Ax_i +BF\zeta_i
\end{align*}
with $i=1, \dots, N$. Then, the overall dynamics of the $N$ agents can be written as
\begin{align}\label{interc_x_proof}
    \dot{x}=(I_N \otimes A +  \mathcal{L} \otimes BF)x.
\end{align}
It has been shown in~\cite{165},~\cite{166}, that the synchronization of the system~\eqref{interc_x_proof} is equivalent to the asymptotic stability of the following $N-1$ subsystems
\begin{align}\label{subsyst_sync_aux}
    \dot{\tilde{\eta}}_i=(A+\lambda_i BF)\tilde{\eta}_i, \hspace{7mm} i=2,\dots, N
\end{align}
where $\lambda_i$, $i=2,\dots, N$ are the nonzero eigenvalues of $\mathcal{L}$.
\begin{lem}[Theorem 2.5~\cite{book_ref}]\label{aux_mainthm}
    The MAS~\eqref{MAS_2.1} achieves state synchronization 
    if and only if the subsystems~\eqref{subsyst_sync_aux}
are globally asymptotically stable.
\end{lem}
\begin{proof}
$(\Longleftarrow)$: Note that $\mathcal{L}$ has eigenvalue 0 with associated eigenvector $\mathds{1}$. Let 
\begin{align*}
    \mathcal{L}=TJT^{-1}
\end{align*}
where $J$ is the Jordan canonical form of the Laplacian matrix $\mathcal{L}$ such that $J(1,1)=0$ and the first column of $T$ equals $\mathds{1}$. Let 
\begin{align*}
    \eta:=(T^{-1} \otimes I_n)x= (x_1, \dots, x_N)^{\top}
\end{align*}
where $\eta_i \in \mathbb{C}^n$. In the new coordinats, the dynamics of $\eta$ can be written as
\begin{align}\label{inter_eta_proof}
    \dot{\eta}=(I_N \otimes A + J_{\mathcal{L}} \otimes BF)\eta.   
\end{align}
By assumption, the subsystems
\begin{align*}
    \dot{\tilde{\eta}}_i=(A+\lambda_i BF)\tilde{\eta}_i, \hspace{7mm} i=2,\dots, N
\end{align*}
are globally asymptotically stable for all $i=2, \dots, N$. Thus, $\eta_i(t)\rightarrow 0$ for $i=2,\dots N$, which implies that
\begin{align*}
    x(t)-(T \otimes I_n)\begin{pmatrix}
\eta_1(t) &
0 &
\cdots &
0
\end{pmatrix}^\top\rightarrow 0.
\end{align*}
Note that the first column of $T$ is equal to the vector $\mathds{1}$. Therefore,
\begin{align*}
    x_i(t)-\eta_1(t) \rightarrow 0
\end{align*}
for $i=1, \dots, N.$ This implies that state synchronization is achieved.

$(\Longrightarrow)$: Suppose that the network~\eqref{interc_x_proof} reaches state synchronization. In this case, we shall have
\begin{align*}
    x(t)-\mathds{1} \otimes x_1(t) \rightarrow 0
\end{align*}
for all initial conditions. Then $\eta(t)-(T^{-1}\mathds{1})\otimes x_1(t) \rightarrow 0$. Since $\mathds{1}$ is the first column of $T$, we have 
\begin{align*}
T^{-1}\mathds{1}=\begin{pmatrix}
1 &
0 &
\cdots &
0
\end{pmatrix}^\top
\end{align*}
Therefore, $\eta(t)-(T^{-1}\mathds{1})\otimes x_1(t) \rightarrow 0$ implies that $\eta_1(t)-x_1(t) \rightarrow 0$ and $\eta_i(t)\rightarrow 0$ for $i=2,\dots, N$ for all initial conditions. This implies that $A+\lambda_i BF$ is Hurwitz stable for $i=2, \dots, N$.
\end{proof}
\begin{rem}
    The function $\eta_1$ defined in Lemma~\ref{aux_mainthm} satisfies
    \begin{align*}
        \dot{\eta}_1=A\eta_1, \hspace{4mm} \eta_1(0)=(w \otimes I_n)x(0).
    \end{align*}
    which can be shown by using the fact that 0 is a simple eigenvalue of the Laplacian. Here, $w$ represents the first row of $T^{-1}$, i.e., the normalized eigenvector associated with the zero eigenvalue, which corresponds to $\mathds{1}$, in the case of an undirected graph. Consequently, the proof of Lemma~\ref{aux_mainthm} shows that the synchronized trajectory of the network is given by,
    \begin{align}\label{sync_traj}
        x_s= e^{At}\frac{1}{N} \sum_{i=1}^N x_i(0).
    \end{align}
\end{rem}
Next, we present the two main sources of inspiration for the presented work: The Algebraic Ricatti equation (ARE)-based protocol presented in~\cite{book_ref} and the Linear Regulator (LR) framework presented in~~\cite{LR_paperII}. 
\subsection{ARE-based Protocol}
In~\cite{book_ref}, a protocol design method based on an algebraic Riccati equation (ARE) is presented. For this protocol, it is shown that full-state coupling is always solvable for a family in $\mathbb{G_{\beta}}$, where no upper bound is required for the nonzero eigenvalues of $\mathcal{L}$.
\begin{prot}[ARE-based Protocol~\cite{book_ref}]\label{protocolARE_2.1}
    Consider a MAS described by~\eqref{MAS_2.1} and~\eqref{zeta_aij}. We consider the protocol
    \begin{align}\label{prot_gral}
        u_i=\rho F \zeta_i,
    \end{align}
    where $\rho \geq 1$ and $F=-B^{\top} P$ with $P>0$ being the unique solution of the continuous-time algebraic Riccati equation
    \begin{align}\label{ARE_LQR}
        A^{\top} P +PA - 2\beta PBB^{\top} P+Q=0
    \end{align}
    where $Q>0$ and $\beta$ is a lower bound for the real part of the nonzero eigenvalues of all Laplacian matrices associated with a graph in the set of connected graphs $\mathbb{G}_\beta$.
\end{prot}
\subsection{Linear Regulator}
An explicit solution for linear positive systems, in the form of a linear algebraic equation, is derived from the problem setting introduced in ~\cite{LR_paperII}. Next, we recall the result that inspired our novel protocol to be presented in Section~\ref{sec/main_res}.
\begin{thm}\label{LR_thm}
    Let $\tilde{A} \in \mathbb{R}^{n\times n}$, $\tilde{B}\in \mathbb{R}^{n\times m }$, $\tilde{E}  \in \mathbb{R}_{+}^{m \times n}$, $s \in \mathbb{R}^{n}_+ $. Suppose that $\tilde{A}- |\tilde{B}| \tilde{E}$ is Metzler. Then the following optimal control problem has a finite value for every $x_0\in \mathbb{R}_+^n$
    \begin{align}\label{gral_min_setting}
    &\underset{\mu}{\inf} \hspace{1mm} \int_{0}^{\infty} \left [ s^{\top}x(\tau) \right ]d\tau \notag \\
            &\mathrm{Subject \hspace{2mm} to} \notag \\
        &~~~~~~~~\dot{x}(t)=\tilde{A}x(t)+\tilde{B}u(t), \hspace{1mm} x(0)=x_{0}\\
        &~~~~~~~~u(t)=\mu(x(t)), \hspace{1mm} \left | u \right | \leq \tilde{E}x. \notag
\end{align}
if and only if there exists a nonnegative vector $p\in \mathbb{R}_{+}^{n}$ such that
\begin{align}\label{ARE_LR}
    \tilde{A}^{\top}p&= \tilde{E}^{\top} |\tilde{B}^{\top}p| - s.  
\end{align}

If either are satisfied, then~\eqref{gral_min_setting}  has the minimal value $p^{\top}x_0$. Moreover, the control law $u(t)=-Kx(t)$ is optimal when
\begin{align} \label{K_LR}
    K := \mathrm{diag}\left(\mathrm{sign}( \tilde{B}^{\top}p)\right)\tilde{E}.
    \end{align}
\end{thm}
\begin{rem}
    The sparsity structure of the control gain $K$ in \eqref{K_LR} is inherited from the $E$ matrix which can be determined by the problem designer. 
\end{rem}
\begin{proof}
    The proof of this theorem relies on dynamic programming theory and, assuming no disturbance enters the local systems, follows from Theorem 2 in ~\cite{LR_paperII}.
\end{proof}

An important notion for this paper is the following.
\begin{defn}[$E$-stabilizability]
    Let $A \in \mathbb{R}^{n\times n}$, $B \in \mathbb{R}^{n\times m }$ and $E  \in \mathbb{R}_{+}^{m \times n}$. We say that a pair $(A,B)$ is $E$-stabilizable if there exists a feedback law $u=-Kx$ with $\left|u \right|\leq Ex$ such that $A-BK$ is Hurwitz.
\end{defn}
\begin{rem}
    To verify the $E$-stabilizability of the pair $(A,B)$ with $A$ Metzler, it is shown in \cite[Lem. 4]{LR_paperII} that it is necessary and sufficient to verify the feasibility of
    \begin{align*}
         Ax + Bu \le - \mathds 1, \quad -Ex \le u \le Ex.
    \end{align*}
    This feasibility problem can be verified by any linear program solver.
\end{rem}
\begin{lem}[Cor. 8 and Thm. 9 ~\cite{LR_paperII}]
    Suppose $s>0$ and that the pair $(\tilde A,\tilde B)$ is $\tilde E$-stabilizable. Then~\eqref{ARE_LR} has a solution $p\geq 0$. The vector $p\geq 0$ solves~\eqref{ARE_LR} if and only if $p$ maximizes linear program
    \begin{align}\label{LP}
        &\mathrm{Maximize} \hspace{2mm} \mathds{1}^{\top}p \hspace{1mm} \mathrm{over} \hspace{1mm} p \in \mathbb{R}^{n}_{+}, \hspace{1mm} \zeta \in \mathbb{R}^{m}_{+} \notag\\
        &\mathrm{Subject} \hspace{1mm} \mathrm{to} \hspace{2mm} \tilde{A}^{\top}p \leq \tilde{E}^{\top}\zeta -s  \\
        &\hspace{15mm} -\zeta \leq \tilde{B}^{\top}p \leq \zeta . \notag
    \end{align} 
\end{lem}
\section{Problem Formulation and Main Results}\label{sec/prob_form}\label{sec/main_res}
This paper presents a solution to two problems:
\begin{prob}[Synchronization problem]\label{prob1}
    Design a linear feedback controller of the form~\eqref{prot_gral} that solves Problem~\ref{problem_gral} and satisfies the constraint
    \begin{align}\label{u_bound_gral}
        \left|u_i \right| \leq {E} \left|\zeta_i \right|.
    \end{align}
\end{prob}
\begin{prob}[Positive Synchronization problem]\label{prob2}
    Find necessary and sufficient conditions for our protocol, which ensures that the positivity of each agent's trajectory is preserved for all nonnegative initial conditions.
\end{prob}
\begin{rem}
    For $B$ nonnegative we always have that $K=E$, implying that the internal positivity of each agent ensures the positivity of the interconnected system. 
\end{rem}

The main contributions of this paper are presented: an LR-based protocol design that solves Problem~\ref{prob1} in Section~\ref{sec/prob_form}, and sufficient and necessary conditions for solving Problem~\ref{prob2}. 
\begin{prot}[LR-based protocol]\label{protocol_LR}
    Consider the MAS described by~\eqref{MAS_2.1} and~\eqref{zeta_aij} with $A \in \mathbb{R}^{n\times n}$ Metzler and $B \in \mathbb{R}^{n\times m }$. Let $E \in \mathbb{R}_{+}^{m \times n}$, $s \in \mathbb{R}^{n}_+$ and $\mathcal{L}$ be a Laplacian matrix associated with a graph $\mathcal{G}\in \mathbb{G}_{[\beta, \gamma]}$ with $N$ agents.
    Suppose \begin{align}\label{ass_betas_cont}
        A-\gamma | B | E
    \end{align}
    is Metzler and $\rho \ge \frac{1}{\beta}$.  
    The LR-based protocol is given by
    \begin{align}\label{u_prot_LR}
        u_i=-\rho K \zeta_i,
    \end{align} 
    where $K$ follows from Theorem~\ref{LR_thm} with $\tilde A = A$, $\tilde B = B$, $\tilde E = \frac 1 \rho E$ and $s > 0$.
\end{prot}

\begin{thm}\label{protocol_thm_LR}
    Consider a graph family $\mathcal{F} \subseteq \mathbb{G}_{[\beta, \gamma]}$ and the MAS described by~\eqref{MAS_2.1} and~\eqref{zeta_aij}. If the pair $(A,B)$ is $E$-stabilizable then the protocol~\eqref{u_prot_LR} solves the state synchronization problem for any undirected graph  $\mathcal{G} \in \mathcal{F}$. Moreover, the synchronized trajectory is given by~\eqref{sync_traj}, and each $u_i$ satisfies the bound \eqref{u_bound_gral}.
\end{thm}
\begin{proof} Let $\mathcal{G}$ be any graph in $\mathcal{F}$. The overall dynamics of the $N$ agents with the protocol~\eqref{protocol_LR} can be written as
    \begin{align}\label{interc_x_proof}
        \dot{x}=(I_N \otimes A - \rho \mathcal{L}(\mathcal G) \otimes BK)x.
    \end{align}
By Lemma~\ref{aux_mainthm}, the synchronization of the system~\eqref{interc_x_proof} is equivalent to the asymptotic stability of 
\begin{align}\label{diag_agents}
    \dot{\tilde{\eta}}_i=(A-\lambda_i \rho BK)\tilde{\eta}_i, \hspace{7mm} i=2,\dots, N
\end{align}
where $\lambda_i=\lambda_i(\mathcal{G})$, $i=2, \dots, N$ are positive. We prove that $A-\lambda_i \rho BK$ is Hurwitz for all $i$.

Let $i=2, \dots, N$ and $\alpha_i=\lambda_i \rho$. Recall that $\lambda_i \ge \beta \ge \frac 1 \rho$, and thus $\alpha_i \ge 1$. By assumption, $\gamma \ge \lambda_i$ and $A- \gamma \rho  \left|B \right|\tilde E$ is Metzler and since $BK \leq |B|\tilde E$, $A-\alpha_i BK$ is also Metzler. 
Observe from~\eqref{K_LR} that $K$ satisfies
\begin{align*}
    K^{\top}(B^{\top}p)=\tilde E^{\top}\left| B^{\top}p \right|
    \ge 0,
\end{align*}
where $p \geq 0$ solves~\eqref{ARE_LR}.
It follows from~\eqref{ARE_LR} that
\begin{multline*}
    (A-\alpha_i BK)^{\top}p=A^{\top}p-\alpha_i K^{\top}B^{\top}p\\
    =\tilde E^{\top}\left| B^{\top}p \right| - s - \alpha_i \tilde E^{\top}\left| B^{\top}p\right|\\
        =(1-\alpha_i)E^{\top}\left| B^{\top}p\right| -s 
        \le -s < 0.
\end{multline*}
Therefore, by \cite[Thm. 5.1]{fiedler1986} it is implied that $A-\alpha_i BK$ is Hurwitz as we wanted to prove.
\end{proof}

Positive systems often appear as dynamical systems for which the states are representing (physical) quantities that cannot become negative. In some contexts it is therefore desired that the trajectories to remain in the nonnegative orthant. The following theorem characterizes when this is the case.
\begin{thm}\label{thm_nonneg}
    Consider a graph family $\mathcal{F} \subseteq \mathbb{G}_{[\beta, \gamma]}$ and the MAS described by~\eqref{MAS_2.1} and~\eqref{zeta_aij}. Suppose the pair $(A,B)$ is $E$-stabilizable and consider the protocol~\eqref{u_prot_LR}. The trajectories of the MAS remain nonnegative for all nonnegative initial conditions if and only if $BK$ is nonnegative.
\end{thm}
\begin{proof}
The dynamics~\eqref{MAS_2.1} of each agent $i$ is rewritten \nobreak{as}
\begin{align}\label{MAS_pf_thm7}
    \dot{x}_i=\hat{A}x_i + \hat{B} \hat{u}_i 
\end{align}
where $\hat{A}=A-\rho BK \sum_{j=1}^N a_{ij}$, $\hat{B}=\rho BK$ and $\hat{u}_i=\sum_{j=1}^N a_{ij}x_j$.
Note that $\gamma \geq \sum_{j=1}^N a_{ij}$ and $|B|E \geq BK$, hence, $\hat{A}$ is Metzler.

($\Longrightarrow$):  Suppose that the matrix $BK$ has, at least, one negative element, then there exists $\hat{B}_{p,q} <0$ for some $p,q \in \mathbb{N}$. Let $x_i(0)_p=0$ and $x_j(0)_q$ be sufficiently large for some $q\neq p$. Then $\hat{u}_i(0)_q \gg 0$ and from~\eqref{MAS_pf_thm7} it follows that $\dot x_i(0)_p <0$. Thus, $x_i$ leaves the nonnegative orthant.

($\Longleftarrow$): Because $\hat A$ is Metzler and $\hat B \geq 0$ the system is internally positive with respect to $\hat u$~\cite{tutorial}. Thus, the trajectories of~\eqref{MAS_pf_thm7} remain nonnegative. 
\end{proof}

Recall that the LR-based consensus protocol requires that the gain parameter $\rho$ satisfies $\rho \ge \frac 1 {\lambda_2}$. Violation of this bound may destabilize the systems \eqref{diag_agents} for some $i$, resulting in dissensus. At the same time we require in the proof of Theorem~\ref{protocol_thm_LR} that the state matrix $A-\lambda_i \rho BK$ in \eqref{diag_agents} is Metzler, in order to apply the Linear Regulator Theorem~\ref{LR_thm}. For this to hold, we need $\lambda_i \rho \le \alpha$, where
\begin{align}\label{alpha}
    \alpha = \operatorname{argmax}_{\tau \ge 0}\left\{A- \tau |B|\tilde E \text{ is Metzler} \right\}, \tilde E = \tfrac 1 \rho E.
\end{align}
For any graph family $\mathcal{F}\subset\mathbb{G}_{[\beta,\gamma]}$ we have $\lambda_i(\mathcal G)\in [\beta,\gamma]$ for any $\mathcal{G}\in\mathcal{F}$, implying the bounds
\begin{align*}
    \tfrac 1 {\lambda_i} \le \tfrac 1 \beta \le \rho \le \tfrac \alpha \gamma \le \tfrac \alpha {\lambda_i}.
\end{align*}
We therefore conclude that, in order to apply the LR-based protocol, we require that
\begin{align}\label{alpha_condition}
    \alpha \ge \frac \gamma \beta .
\end{align}
Note that $\frac \gamma \beta \ge 1$, and thus $\alpha\ge 1$ is required as well.
Note also that the left-hand side of \eqref{alpha_condition} depends fully on the local agent dynamics via \eqref{alpha}, whereas the right-hand side depends on the class $\mathbb{G}_{[\beta,\gamma]}$ that contains the graph family of interest. 
It follows that the system dynamics dictate for which graph families the Protocol~\ref{protocol_LR} can reach consensus.
Although the matrix $\tilde E$ could be scaled down such that $\alpha$ is increased, doing so might violate the $E$-stabilizability of $(A,B)$.


\section{Graph Families with Eigenvalue Bounds}\label{sec/graphs}
The Linear Regulator-based protocol proposed in Protocol~\ref{protocol_LR} is suitable for the application to graph families for which upper and lower bounds on the Laplacian eigenvalues are available.
This approach can therefore be of interest for problems where the graph topology is uncertain, but general bounds on graph theoretical notions such as the maximum degree, the average degree, the graph diameter, or even the Cheeger constant or the chromatic number are known.

The literature on graphs and their spectral properties is vast and diverse, and providing a thorough discussion is beyond the scope of this paper. 
Instead we will present several points of reference that might be insightful and relevant to the presented protocol.
The classical paper \cite{anderson1985eigenvalues} establishes the simple bound 
$\lambda_N \le \max_{(i,j)\in\mathcal{E}(\mathcal G)}(d_i+d_j)$, where $d_v$ is the degree of node $v$.
For $d$-regular graphs this implies that the maximal eigenvalue $\lambda_N$ is upper bounded by $2d$.
The paper \cite{LIN201611} presents upper and lower bounds on the Laplacian upper and lower bounds for the nonzero eigenvalues which are tight for strongly regular graphs.
These bounds can be used to establish an upper bound for the ratio $\frac \gamma \beta$ in terms of the number of nodes, the degrees of the agents, and the overlap of neighbors of each pair of nodes.
We refer to \cite[Sec. 2.7]{stanic2017regular} for general bounds on the nonzero eigenvalues of Laplacian matrices.

Expender graph families are characterized by the property that the algebraic connectivity $\lambda_2$ of the graphs satisfies $0<\beta<\lambda_2$.
Other graphs such as small-world networks model may also be considered. Well-known examples are the Barab\'asi–Albert model, the Erd\H{o}s–R\'enyi model, and the Watts–Strogatz model.
The paper \cite{mirchev2017spectra} illustrates how the Laplacian eigenvalues of graphs in these families are typically distributed.

\section{Simulations}
Consider the MAS~\eqref{MAS_2.1} described by 
\begin{align*}
    A= \begin{bmatrix}
        -2.21 &  2.40\\
        0.43 & -0.44
    \end{bmatrix}; \hspace{3mm} B = \begin{bmatrix}
        0.27 \\
        0
    \end{bmatrix};\hspace{3mm}
    E=\begin{bmatrix}
        0.06\\ 
        0.6
    \end{bmatrix}^{\top}
\end{align*}
and composed by 150 agents. Consider also a connected undirected graph $\mathcal{G}$ in the family of regular graphs of degree $d=5, 7$ for $\mathcal{F}_R$. The matrix A is unstable $\sigma(A)=\left\{-2.63,  0.03\right\}$.

Let the eigenvalues of every $\mathcal G \subseteq \mathcal F_R$ be upper bounded by $\gamma=13$ and lower bounded by $\beta = 1$ such that $\rho=1/ \beta=1$. To solve the state synchronization problem, the LR-based Protocol~\ref{protocol_LR} is implemented. Consider $s=\mathds 1>0$, and $\tilde E =\frac{1}{\rho} E =\begin{bmatrix}
    0.06 & 0.6
\end{bmatrix}$ such that $A- |B|\tilde E$ is Metzler. The linear program~\eqref{LP} is maximized by a vector $p^*$, which satisfies the algebraic equation~\eqref{ARE_LR} resulting in $K=E$. 

Observe that, $A-\gamma |B|E$ with $\tilde E = \frac 1 \rho E$ is Metzler and $BK=BE \geq 0$. Hence, from Theorem~\ref{thm_nonneg} the trajectories of the MAS remain nonnegative for all initial conditions. The initial condition is arbitrarily chosen to be nonnegative.

\begin{figure}[!t]
    \centering
    \includegraphics[scale=0.3]{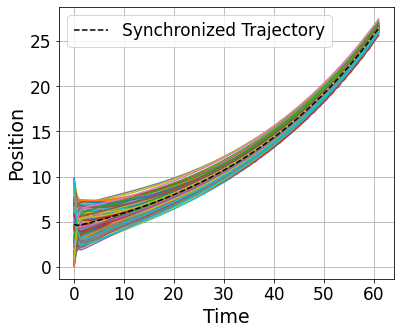}
    \includegraphics[scale=0.3]{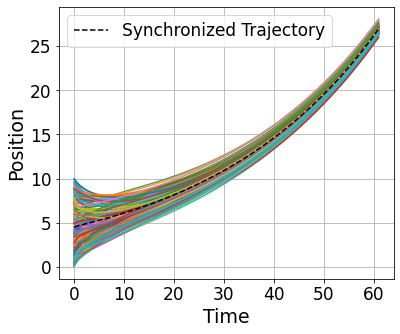}
    \includegraphics[scale=0.3]{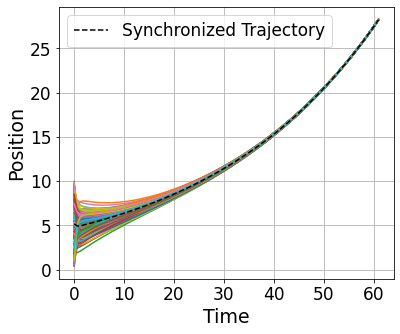}
    \includegraphics[scale=0.3]{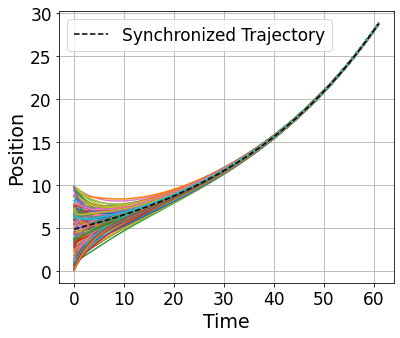}
    \caption{Evolution over time of the first (left panels) and the second (right panels) state of each agent $i=1, \dots, 150$ synchronizing over $5$-regular graphs (upper panels) and $7$-regular graphs.}
    \label{fig_sync}
\end{figure}

\begin{figure}[!t]
    \centering
    \includegraphics[scale=0.25]{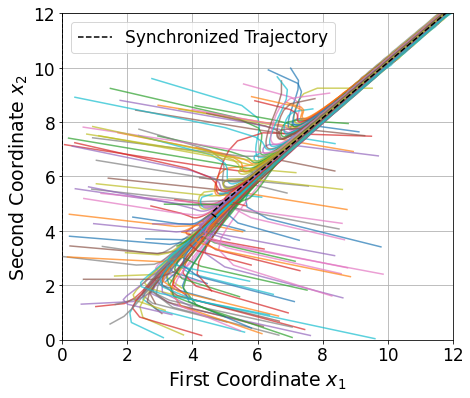}
    \includegraphics[scale=0.25]{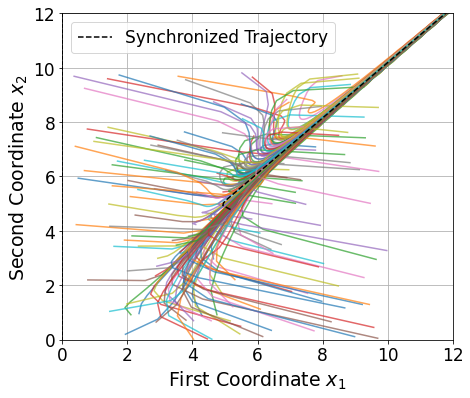}
    \caption{Trajectories of agents synchronizing over a $5$-regular graph (left panel) and $7$-regular graph.}
    \label{fig_2d}
\end{figure}

Figure~\ref{fig_sync} illustrates the evolution of the first and second states of each of the 150 agents in an interconnected system, where each agent is connected to 5 and 7 neighbors, respectively. In Figure~\ref{fig_2d} the state synchronization of the agents is illustrated with a 2D plot of their trajectories. As expected, the trajectories do not leave the positive orthant. Moreover, the synchronization is achieved faster as the nodal degree is increased. 

\section{Conclusions}
This paper introduced the linear-regulator consensus protocol for multi-agent systems with positive dynamics.
The protocol was presented in the context of graph families, with the goal to achieve synchronization in all graphs within the family.
It was assumed that these graphs upper and lower bounds on the eigenvalues of the graphs are known.
The approach may be regarded as a positive systems-analogue of the well-established LQR-based consensus protocol. 
Based on recent results on the Linear Regulator problem, it was shown that the presented method achieves stability, and that for each agent the control law satisfies an \textit{a priori} proportional bound on the input.
Moreover, a condition for the positivity of all state trajectories was given.
A linear program was presented to find the optimal $p$ from which the control is derived. 
%
Given the abundance of application of positive system, we believe that the proposed protocol can be a valuable contribution for addressing synchronization problems in real world systems.

Future work includes the extensions of the protocol to discrete time and to directed graphs. Another extension is to derive a priori conditions that guarantee positive synchronization. 
Finally, a more extensive treatment of different graph families is necessary, with a focus on the performance of the protocol.  

\bibliographystyle{unsrt} 
\bibliography{bibliography}
\end{document}